\documentclass[a4paper]{jpconf}
\usepackage{graphicx}
\begin{document}
\title{Dark Matter Candidates:  What Cold, ..and What's Not}

\author{Lawrence M. Krauss}

\address{CERCA, Physics Department, Case Western Reserve University, 10900 Euclid Ave, Cleveland OH 44106-7079, and Physics Department, Vanderbilt University, Nashville TN.}

\ead{krauss@cwru.edu}

\begin{abstract}
In this brief review of recent theoretical developments associated with the search for dark matter I describe the following: why baryons are now ruled out as dark matter candidates; SUSY WIMPS and signatures in the MSSM and NMSSM why claimed indirect signatures are probably not WIMP related, why axions may be of new interest, how WIMP detection might tell us about the galactic halo, and how theorists are preparing to avoid the next generation of experimental constraints. (Invited review talk, Neutrino 2006, Santa Fe 2006) 
\end{abstract}

\section{Introduction}
Determining the nature and origin of the dark matter dominating almost all clustered systems in the universe remains, some 40 years after strong evidence of the existence of dark matter was first presented, one of most important outstanding questions in physics and cosmology. While light neutrinos were the first natural non-baryonic candidate for dark matter, there is now ample evidence that this material, while definitely non-baryonic as I shall describe, must also be 'cold', i.e.  non-relativistic at the time the first large scale structures in the universe began to form.  Perhaps the favored Cold Dark Matter candidate is the Lightest Supersymmetric Particle (LSP), which not only arises naturally in extensions of the Standard Model, but also has a mass scale and interaction strength (making it a WIMP-Weakly Interacting Massive Particle) which naturally falls in the range allowing it to possibly dominate the mass density of the universe today.  Nevertheless, as collider experiments continue to constrain the scale of SUSY breaking, the LSP allowed range is shrinking, and questions of fine tuning have arisen.   For this reason, as I shall describe, extensions of the Minimal Supersymmetric Standard Model are now being considered.   For all such WIMPS, there has been potential excitement associated with possible indirect signatures from annihilation in the galaxy, and I review why such excitement is misplaced.   At the same time, as the second generation of WIMP detectors is coming online and beginning to seriously probe the LSP parameter space, it is worth considering how the uncertainties associated with our lack of knowledge of our galactic halo might impact upon such experiments, and also how one might design experiments that can separate astrophysical uncertainties from particle physics uncertainties.   Next, as I shall describe, axions, the 'other' well motivated CDM candidate are once again returning from obscurity, as particle physicists are finding new ways of making unnatural acts natural.  Finally, of course, there are a host of unmotivated dark matter candidates that are being discussed, demonstrating once again that beauty, even in science, is in the eye of the beholder.

\section{Baryons Aren't }

From the moment that Dark Matter was first inferred, using the motion of galaxies in clusters, and then, more solidly, by the rotation curves of spiral galaxies, the natural suspicion was that it could easily be explained as being due to non-luminous baryons, like snowballs, planets, or brown dwarfs.  Over time these possibilities became more and more constrained.  By the 1990's, one of the most severe constraints came from Big Bang Nucleosynthesis (BBN).  

Observations of absorption of light from distant quasars by intervening hydrogen gas clouds allowed for the first time a measurement of the fraction of primordial deuterium in these clouds.  
While the measurements are quite difficult, and subject to large possible systematic uncertainties, the measured deuterium to hydrogen fraction settled on a value of approximately $ 3 \times 10^{-5}$.  Comparing this to the predictions of standard BBN calculations yields a baryon fraction (compared to the critical density) of $ \Omega_B h^2 = .02 \pm .002$, where $h$ represents the Hubble constant in units of $100 km/s/Mpc$, and HST values suggest $h \approx 0.7$.   

At the same time, separate astrophysical observations, ranging from X-Ray studies of galaxy clusters to gravitational lensing observations of these system began to converge on an inferred dark matter fraction that was statistically incompatible with this fraction, in the range $\Omega h^2 = 0.1-0.15$.  

While this inconsistency was apparent, there were still loopholes.  First, could BBN estimates be trusted?  Second, large scale structure observations had often been subject to large systematic uncertainties that had caused cosmologists to refine estimates of $\Omega_{clustered}$ on numerous occasions.  

Happily, whatever nagging doubts may have existed have largely been laid to rest following the WMAP observations of the CMB, which have independently confirmed both of the above estimates for $\Omega_B$ and $\Omega_{clustered}$ \cite{WMAP}.   As a result, we now have definitive evidence that dark matter {\it  cannot} be baryonic, and therefore is likely composed from a gas of more exotic elementary particles.

\section{The Usual Suspects}

I like to categorize non-baryonic dark matter candidates by the different mechanisms by which they may have arisen.   In this case, dark matter candidates fall into one of the following categories:

\begin{itemize}
\item {\bf BORN TO BE DARK},
\item {\bf ACHIEVE DARK MATTERDOM}
\item {\bf HAVE DARK MATTERDOM THRUST UPON THEM}
\end{itemize}

\subsection{ Born to be Dark}

The prototypical such candidate is a light neutrino.  Present in roughly the same thermal equilibrium numbers in the early universe as photons, which have an energy fraction of roughly $\Omega \approx 10^{-5}$, if $m_{\nu}/T \approx  10^5$ then neutrinos will automatically close the universe.  Unfortunately they don't.  Or rather, fortunately, they don't, because if they did it is not clear that galaxies would have formed in time for us to be here today.

\subsection{Achieve Dark Matterdom}

Here, WIMPS are the prototypical candidate.  Like neutrinos, they were present with thermal equilibrium densitys comparable to photons at early times.  However, their annihilation cross-sections, which scale with their mass, are much larger.  As a result, before their interactions freeze out, the temperature of the universe would have decreased below their mass.  As a result, their number density will be suppressed by a factor of roughly $ exp[-M/T_{freeezeout}]$ compared to photons.  This produces a roughly critical mass today if $M \approx O(GeV)$ and $ exp[-M/T_{freeezeout}] \approx 1/20$, which requires interaction cross sections that are roughly comparable to weak interaction cross sections.

\subsection{Have Dark Matterdom Thrust Upon Them}

Axions are the prototypical candidate in this regard.   With predicted masses much less than 1 $ eV$, axions by all rights should be cosmologically irrelevant.  However, cosmic axions are not be produced just thermally.  Since axions are the pseudo-Goldstone boson associated with a global phase transition, they can exist as a Bose condensate at early times.  If, for example, the Peccei-Quinn phase transition happens before Inflation, the axion field, described by an angular variable, $\theta=a/F$, where $F$ is the PQ symmetry breaking scale, can have a non-zero expectation value throughout the entire universe.   While the potential is strictly flat, when QCD effects break the symmetry, the axion field gets a mass.  This induced curvature implies that there is non-zero energy density stored in the coherent axion field.  It is miniscule at early times, but if the axion mass is very small, the axion field does not begin to relax down the potential until late times.  Until it relaxes, the energy density stored in the field looks like a cosmological term, and remains constant, while the matter density of the rest of the universe falls as $R^{-3}$.  Once the field starts to relax, the coherent field energy gets converted into non-relativistic matter, whose energy begins to redshift. Ultimately, if the initial value of $\theta$ is $O(1)$ then the axion can dominate the energy density of the universe, with a contribution to $\Omega$ of

\begin{equation}
\Omega_a \approx (10^{-5} ev/m_a) (200 MeV/\Lambda_{QCD})
\end{equation}

\section{Problems in SUSY Paradise: SUSY Dark Matter and the NMSSM}

The Minimal Supersymmetric Standard Model (MSSM) provides an elegant framework in which to attempt to understand two central issues in elementary particle physics, the hierarchy problem, and the possibility of Grand Unification.   It is remarkable that at the same time the energy scale associated with low energy supersymmetry breaking can naturally result in stable particles whose interaction strength is precisely in the range described above as required for WIMPs.

And if nature were an impressionist painting the MSSM would, without question, be considered the most likely candidate for physics beyond the standard model.   However, when examined in detail, certain issues arise which suggest some fine tuning might be necessary.  These are:

\begin{itemize}
\item{ THe $\mu$ problem:  A SUSY conserving mass term for the two higgs superfields in the MSSM with mass parameter $\mu$ can have values ranging from zero to the Planck Scale.  Why it should be fine-tuned at the electroweak scale is not clear}
\item{Non observation of the Higgs:  The upper bound on a neutral Higgs in the MSSM at LEP requires that the Higgs exist very close to its upper bound, requiring fine tuning}
\item{How to get $\Omega <<1$?:  Because annihilation of LSP's determines their remnant abundance, in order to get sufficient annihilation for relatively heavy LSP's, one has to have light intermediate annihilation channels.  But collider constraints on sleptons and squarks, as well as the Higgs particles make this increasingly difficult.}
\item{small flavor changing rates require fine tuning}
\item{parameter space squeezed: the parameter space of the MSSM is being increasingly squeezed, especially if dark matter LSP's is desired}
\end{itemize}

Within the context of the MSSM, a number of authors have examined what the implications of existing collider constraints are for dark matter searches.   It has been claimed that constraints on $\Omega$, combined with collider constraints drive the allowed range of models to be those with small $\mu$ parameter.  In some cases this gives a lower bound on the WIMP spin-independent cross section with nucleons, the parameter of relevance to direct dark matter detectors \cite{kitano}.  It has also been claimed \cite{carena} that Tevatron searches for the neutral Higgs, which are sensitive to large $ tan \beta$ and small $m_A$ are precisely the parameter range probed by dark matter experiments like the CDMS experiment \cite{CDMS}.  Thus, direct dark matter constraints will impact on what is observable at the Tevatron.

As an alternative to this possibility, a number of authors have examined a next-to-minimal version of the SSM, in which a gauge singlet Higgs superfield is added.  If this superfield gets a VEV of the order the the SUSY breaking scale, it leads to an effective $\mu$ parameter that is also of the order of the EW scale.  It also allows the upper bound on the neutral Higgs to be increased, and allows for a very light Higgs boson which is not experimentally excluded, and which also provides an additional annihilation channel for SUSY WIMPs, increasing the allowed parameter space as a function of LSP mass \cite{belanger,gunion}.  In particular, very light LSP dark matter masses are now possible.

Recently, my collaborators and I carried out a comprehensive examination of the possibility that these additional phenomenological attractions in the NMSSM might lead to new indirect signatures for the detection of SUSY dark matter.   We have found that extra one-loop amplitudes for NMSSM annihilation into photons and gluons is enhanced, and that low mass antimatter experiments should be a good probe of such NMSSM WIMPs, and that the possibility of the detection of a monochromatic gamma-ray line within the NMSSM is more promising than the MSSM \cite{ferrer}.  It is also possible that NMSSM WIMPs might form an additional solar system dark matter contribution which could enhance detection of annihilation in the Earth \cite{damour}.

\section{Looking for Dark Matter in All the Wrong Places?}

While the direct and indirect signatures for SUSY WIMPs in the MSSM and NMSSM are exciting, motivation for considering the added flexibility of light WIMPs allowed by the NMSSM was provided by several claimed direct and indirect hints of halo dark matter in experiments.   The DAMA experiment, for example, claimed to observe an annual modulation signal in excess of noise in their Sodium Iodide scintillation experiment.  This, however, was inconsistent with limits from the original CDMS experiment unless the WIMP mass was very small.  However, this rationale turns out to have been misplaced, as it now appears that the DAMA experiment, which has its own consistency problems, appears inconsistent improved limits from direct search experiments, even for low mass WIMPs. 

At the same time, positron annihilation signals have been observed from the Galactic Core \cite{integral,knod}, which was thought might be possibly due to annihation of very light WIMPs in the sub-GeV range. 

It turns out however that detailed analyses have shown that WIMP annihilation cannot account for this signature.  In particular, any mechanism which produces energetic positrons will also be accompanied by internal bremsstrahlung photon emission, and if the positrons are created with an energy greater than 20 MeV, this will violate the COMPTEL/EGRET constraints \cite{beac1}.  Moreover, if positrons are produced a mildly relativistic energies, then higher energy gamma rays will be produced due to in-flight annihilations, requiring that the positrons must be injected with $E <3MeV$ \cite{beac2}.  

At the opposite SUSY extreme of very high WIMP masses, a claimed signature of dark matter annihilation came from claims  of an excess in high energy gamma rays in the 100 Gev-TeV range.  However, a careful analysis of the energy spectrum expected from such annihilations does not match the observed flux \cite{zahar} .   

Thus, for the moment at least, it appears that there is as of yet no compelling direct or indirect evidence for signatures for SUSY WIMPs, and that the next generation of direct and indirect detectors, searching for the signals described in the last section, provide our best bet of constraining the SUSY parameter space in a way that will complement the upcoming searches at the LHC.

\section{Using Halo Uncertainties to distinguish Dark Matter from Noise}

If direct search experiments ever do detect a signal, it will in fact resemble noise.  Indeed this fact is one of the reasons that the DAMA claimed detection is so difficult to interpret.   Clearly it will be necessary to consider a second generation of experiments that will be more sensitive to the halo properties of the dark matter, in particular the fact that the Earth and Sun are moving with respect to the galactic rest frame.   A detector with full directional information would be optimal in order to distinguish a preferred direction for nuclear recoils from WIMP interactions.   There are, however, no detectors with such sensitivity.  Happily, we have recently explored the efficacy of using detectors with a two dimensional directional sensitivity, as may be achieved by the proposed DRIFT experiment.   We have shown, as can be seen in the table below \cite{copi}, that {\it if and only if }forward backward sensitivity is possible, i.e. the head of the recoil track can be distinguished from the tail, that 2D detectors which can rotate in the laboratory frame are almost as efficient as full 3D detectors for distinguishing motion through an isothermal halo from a flat laboratory background. 

\begin{center}
\begin{table}[h]
\caption{The number of events required to identify a WIMP signal above a
    flat background for different types of detectors and a WIMP mass of
    $m_\chi=100 GeV$.}
\centering
     \begin{tabular}{lccc}
      \multicolumn{1}{c}{Detector} &
      \multicolumn{3}{c}{$v_0\;\mathrm{(km/s)}$} \\ \cline{2-4}
      \multicolumn{1}{c}{Type} & $170$ & $220$ & $270$ \\ \hline
      3D (full) & $6$ & $11$ & $18$ \\
      3D without FB & $176$ & $1795$ & $>35,000$ \\
      2D---best/worst & $19/45$ & $34/75$ & $61/123$ \\
      2D rotating & $13$ & $24$ & $43 $
    \end{tabular}
\end{table}
\end{center}

\section{Axions are Back?}

Axions, while by far the most elegant solution of the strong CP problem, have been less favored of late as dark matter candidates because the parameter range for allowed axion masses does naturally lie in the range in which axions would be dark matter.  If the Peccei-Quinn scale is near the GUT scale then if $\theta \approx 1$, axions would close the universe today.  Moreover, constraints from axion emission by supernovae, red giants and white drwarfs put a limit on axion masses of less than $O(10^{-3} eV$, so that axion masses are being squeezed from the high end as well.    

However, it is true that if $< \theta> << 1$ in our universe then GUT scale axions could be dark matter.  Until recently this possibility was viewed as unnatural.  However, recently, due to the inability to naturally explain what appears to be a cosmological constant dominating the energy density of the Universe with an absurdly small and non-zero value, theorists have been driven to the last refuge of scoundrels, namely the anthropic principle.   

While much of the discussion regarding anthropics is tantamount to metaphysics, it is true that if inflation occurs after the PQ transition, then the value of $\theta$ will be a random variable over different causally disconnected regions.  Recently it has been argued that if one is to average over universes with sufficient clustered matter, then the expected value of $\theta$ that is favored is such that the axion dark matter density would be comparable to the observed density of dark matter today \cite{tegmark}.  This is amusing, but like all anthropic arguments, far from compelling.  Nevertheless, it has boosted axion stock on futures markets around the world. 

\section{Conclusions:  From the Sublime to the Ridiculous}

Light SUSY WIMPs and axions remain as highly motivated and potentially detectable dark matter candidates.  The possibility of future discoveries in direct and indirect dark matter searches can complement the range accessible at terrestrial accelerators, meaning that the beautiful complementarity between non-accelerator and accelerator physics continues.   Of course, as mentioned, beauty is in the eye of the beholder, and the possibility of dark matter that might arise naturally in particle physics has not stopped theorists from imagining a host of dark matter particles that  are both undetectable and unmotivated.   I see no good reason to review these possibilities here. 
 
\vskip 0.2in 

\noindent {\bf Acknowledgments}
I thank the organizers of Neutrino 2006 for producing a very interesting meeting, and my collaborators, in particular Francesc Ferrer and Craig Copi, for their important contributions to our projects, and for educating me about many things.  My research is supported in part by DOE and NASA grants.

\smallskip

\end{document}